\documentclass[12pt,letterpaper]{JHEP3}
\usepackage{cite}
\usepackage{epsfig}

\title{Precision cosmology and the landscape}

\author{%
Raphael Bousso\\ 
Center for Theoretical Physics, Department of Physics\\
University of California, Berkeley, CA 94720-7300, U.S.A.\\
{\em and}\\
Lawrence Berkeley National Laboratory, 
Berkeley, CA 94720-8162, U.S.A.}

\abstract{% 
  After reviewing the cosmological constant problem---why is $\Lambda$
  not huge?---I outline the two basic approaches that had emerged by
  the late 1980s, and note that each made a clear prediction.
  Precision cosmological experiments now indicate that the
  cosmological constant is nonzero.  This result strongly favors the
  environmental approach, in which vacuum energy can vary discretely
  among widely separated regions in the universe.  The need to explain
  this variation from first principles constitutes an observational
  constraint on fundamental theory.  I review arguments that string
  theory satisfies this constraint, as it contains a dense discretuum
  of metastable vacua.  The enormous landscape of vacua calls for
  novel, statistical methods of deriving predictions, and it prompts
  us to reexamine our description of spacetime on the largest scales.
  I discuss the effects of cosmological dynamics, and I speculate that
  weighting vacua by their entropy production may allow for prior-free
  predictions that do not resort to explicitly anthropic arguments.
}

\preprint{\hepth{0610211}}

\begin{document}

\section{Introduction}

The quest for quantum gravity is driven by a desire for consistency
and unity of physical law.  Quantum mechanics and the general theory
of relativity are hard to fit under one roof.  String theory succeeds
at this task, exhibiting a level of mathematical rigor and richness of
structure that has yet to be matched by other approaches.  

Unfortunately, the subject has been lacking guidance from experiment.
Particle accelerators, in particular, are unlikely to probe effects of
quantum gravity directly.  The energies that can be attained are many
orders of magnitude too low.  This problem has nothing to do with
string theory.  It arises the minute we turn our attention to quantum
gravity, because gravity is extremely weak in scattering experiments.

On large scales, however, gravity rules.  The expansion of the
universe dominates over all other dynamics at distances above 100 Mpc.
Similarly, once matter condenses enough to form a black hole, no known
force can prevent its total collapse into a singularity.  In the early
universe, moreover, quantum effects can be important.  Perhaps, then,
string theory should be looking towards cosmology for guidance.

In fact, recent years have seen a remarkable transformation.  String
theory has become driven, to a significant extent, by the results of
precision experiments in cosmology.  The discovery of dark
energy~\cite{Per98,Rie98} suggests that vacuum energy is an
environmental variable.  String theory naturally provides for
variability of the cosmological constant, with a fine enough spacing
to accommodate the observed value~\cite{BP}.  In this sense, recent
cosmological observations constitute observational evidence for the
theory.  Moreover, they have focussed attention on the large number of
metastable vacua---the string
landscape~\cite{BP,KKLT,Sus03}---believed to be responsible for this
variability.\footnote{For a detailed review and extensive references,
  see Ref.~\cite{DouKac06}.  For a less technical discussion of the
  issues covered in the present article, see Ref.~\cite{BouPol04}.}

Before presenting conclusions from precision experiments, I will argue
in Sec.~\ref{sec-why} that much can be learned from far more primitive
observations of the cosmos.  A simple question---why is the universe
so large?---translates into a number of major challenges to
theoretical cosmology.  One, the flatness problem, motivated the
theory of inflation, which went on to explain the origin of structure
in the universe, making a number of specific predictions.  Another,
the cosmological constant problem, is especially closely related to
fundamental theory: Why is the energy of empty space more than 120
orders of magnitude smaller than predicted by quantum field theory?

Most early discussions of the cosmological constant problem tended to
embrace one of two distinct approaches.  Either the cosmological
constant has to be zero due to some unknown symmetry; or it is an
environmental variable that can vary over distances large compared to
the visible universe, and observers can only live in regions where it
is anomalously small.  Though neither of these approaches had been
developed into concrete models, each made a signature prediction: that
the cosmological constant is zero, or that it is small but nonzero.

The refined experiments of the last ten years have amassed additional
evidence for inflation, and they have managed to discriminate clearly
between the two approaches to the cosmological constant problem.  The
discovery of nonzero dark energy, in particular, is precisely what the
second, environmental approach predicted, and it all but rules out the
first approach.  These results, summarized in Sec.~\ref{sec-pc}, are
the empirical foundation of the landscape of string theory.

In Sec.~\ref{sec-bp}, I will describe a concrete model that realizes
the second approach in string theory.  The topological complexity of
compact extra dimensions leads to an exponentially large potential
landscape.  Its metastable vacua form a dense ``discretuum'' of values
of the cosmological constant.  Every vacuum will be realized in
separate regions, each bigger than the visible universe, but structure
(and thus observers) form only in those regions where the cosmological
constant is sufficiently small.

In Sec.~\ref{sec-predict}, I will discuss some of the novel challenges
posed by the string landscape.  The greatest challenge, perhaps, is to
develop methods for making predictions in a theory with $10^{500}$
metastable vacua.  In fact, this difficulty is sometimes presented as
insurmountable, but I will argue that it just comes down to a lot of
hard work.  In particular, I will argue that the correct statistical
treatment of vacua necessitates a departure from the traditional,
global description of spacetime.  I will further propose a statistical
weighting of vacua based on entropy production, which performs well in
comparison with far more specific anthropic conditions.  A general
weighting of this type may pave the way for a calculation of the size
of the universe from first principles.

\section{Why is the universe large?}
\label{sec-why}

In cosmology, the most naive questions can be the most profound.  A
famous example is Olbers' paradox: Why is the sky dark at night?  In
this spirit, let us ask why the universe is large.  To quantify
``large'', recall that only a single length scale can be constructed
from the known constants of nature: the Planck length
\begin{equation}
  l_{\rm P} = \sqrt{\frac{G\hbar}{c^3}}
  \approx 1.616 \times 10^{-33} {\rm cm}~.
\end{equation}
Here $G$ denotes Newton's constant and $c$ is the speed of
light.\footnote{In the remainder I will work mostly in Planck
  units.  For example, $t_{\rm P}=l_{\rm P} /c \approx .539 \times
  10^{-43} {\rm s}$ and $M_{\rm P} = 2.177\times 10^{-5} {\rm g}$.}

The actual size of the universe is larger than this fundamental length
by a factor 
\begin{equation}
H^{-1} = .8 \times 10^{61}~.  
\label{eq-h}
\end{equation}
Here $H\approx 70$ km/s/Mpc is the Hubble scale, and $H^{-1}$ is the
Hubble length.  Of course, this refers to the size of the universe as
we see it today, and thus is only a lower bound on the length scales
that may characterize the universe as a whole.\footnote{As I shall
  discuss below, there is evidence that the universe is exponentially
  larger than the visible universe, but that we will never see a
  region larger than $.98\times 10^{61}$, no matter how long we
  wait.}

The dynamical behavior of a system usually reflects the scales of the
input parameters, and other scales constructed from them by
dimensional analysis.  For example, the ground state of a harmonic
oscillator of mass $m$ and frequency $\omega$ has a position
uncertainty of order $(m\omega)^{-1/2}$.  The parameters entering
cosmology are $G$, $\hbar$, and $c$, so $l_{\rm P}$ is the natural
length scale obtained by dimensional analysis.  Thus, Eq.~(\ref{eq-h})
represents an enormous hierarchy of scales.  Where does this large
number come from?

At the very end of this article, I will speculate about the origin of
the number $10^{61}$.  For now, let us simply consider the qualitative
fact that the universe is large compared to the Planck scale---a fact
that is plain to the naked eye, no precision experiments required.  We
will see that some of the most famous problems in theoretical
cosmology are tied to this basic observation: the flatness problem,
and the cosmological constant problem.

\subsection{The flatness problem and inflation}
\label{sec-inflation}

We live in a universe that is spatially isotropic and homogeneous on
sufficiently large scales.  The spatial curvature is constant, and it
is remarkably small.  By the Einstein equation, this can be related to
the statement that the average density $\rho$ is not far from the
critical density,
\begin{equation}
\Omega\equiv \frac{\rho}{ \rho_{\rm c}} \sim O(1)~,
\end{equation}
where
\begin{equation}
\rho_{\rm c}= \frac{3H^2}{8\pi} ~.
\end{equation}

This is surprising because it means that the early universe was flat
to fantastic accuracy.  Through much of the history of the universe,
$\Omega$ has been pushed away from $1$.  Einstein's equation implies
that
\begin{equation}
  |\Omega-1| = (\dot{a})^{-2}~,
\label{eq-omo}
\end{equation}
where $\dot a$ is the time derivative of the scale factor of the
universe.  The early universe was dominated by radiation for some
70,000 years, and $a$ was proportional to $t^{1/2}$.  Afterwards it
was dominated by matter for several billion years, with $a\propto
t^{2/3}$.  Curvature would have become dominant ($\Omega\neq O(1)$)
over this time unless
\begin{equation}
|\Omega-1| \lesssim 10^{-59} 
\end{equation}
when the universe began.  This is the flatness problem.

The flatness problem is closely related to our original question:
without flatness, the universe could would not have become large.
Suppose, for example, that the early universe had been tuned to
flatness less precisely, say $\Omega=1+10^{-20}$.  This would have
been a closed universe, which would have expanded to a maximum radius
$10^{20}$ and recollapsed in a big crunch, all within about a time
$10^{20}$.  In other words, this universe would have grown no larger
than a proton and lived for less than $10^{-23}$ seconds.

If the universe had started out slightly underdense (say,
$\Omega=1-10^{-20}$), it would have developed a noticably ``open'',
i.e., hyperbolic spatial geometry after $10^{-23}$ seconds, when the
largest structures were the size of a proton.  After this time,
density perturbations would no longer grow and structure formation
would cease.  The largest coherent structures, each the size of a
proton, would freely stream apart.  There would be no objects
comparable to the size of a planet, let alone galaxies.  In this
sense, the universe would be small.

A solution to the flatness problem appeared in the early 1980's:
inflation.  (It simultaneously addressed a number of other major
conundra, such as the horizon problem.)  For a detailed treatment,
see, e.g., Refs.~\cite{LidLyt93,LidLyt00}.

The idea is to use Eq.~(\ref{eq-omo}) to our advantage: if $\dot a$
increases with time, then $\Omega$ is driven to $1$.  This can be
accomplished by positing that the very early universe was dominated by
the vacuum energy of a scalar field before yielding to the standard
radiation era.  The scale factor grows almost exponentially with time,
and $\Omega$ quickly approaches $1$ with exponential accuracy:
\begin{equation}
|\Omega-1| \approx e^{-2N}~,
\end{equation}
Here $N$ is the number of $e$-foldings, i.e., $e^N$ is the ratio
between the scale factor before and after inflation.
\EPSFIGURE{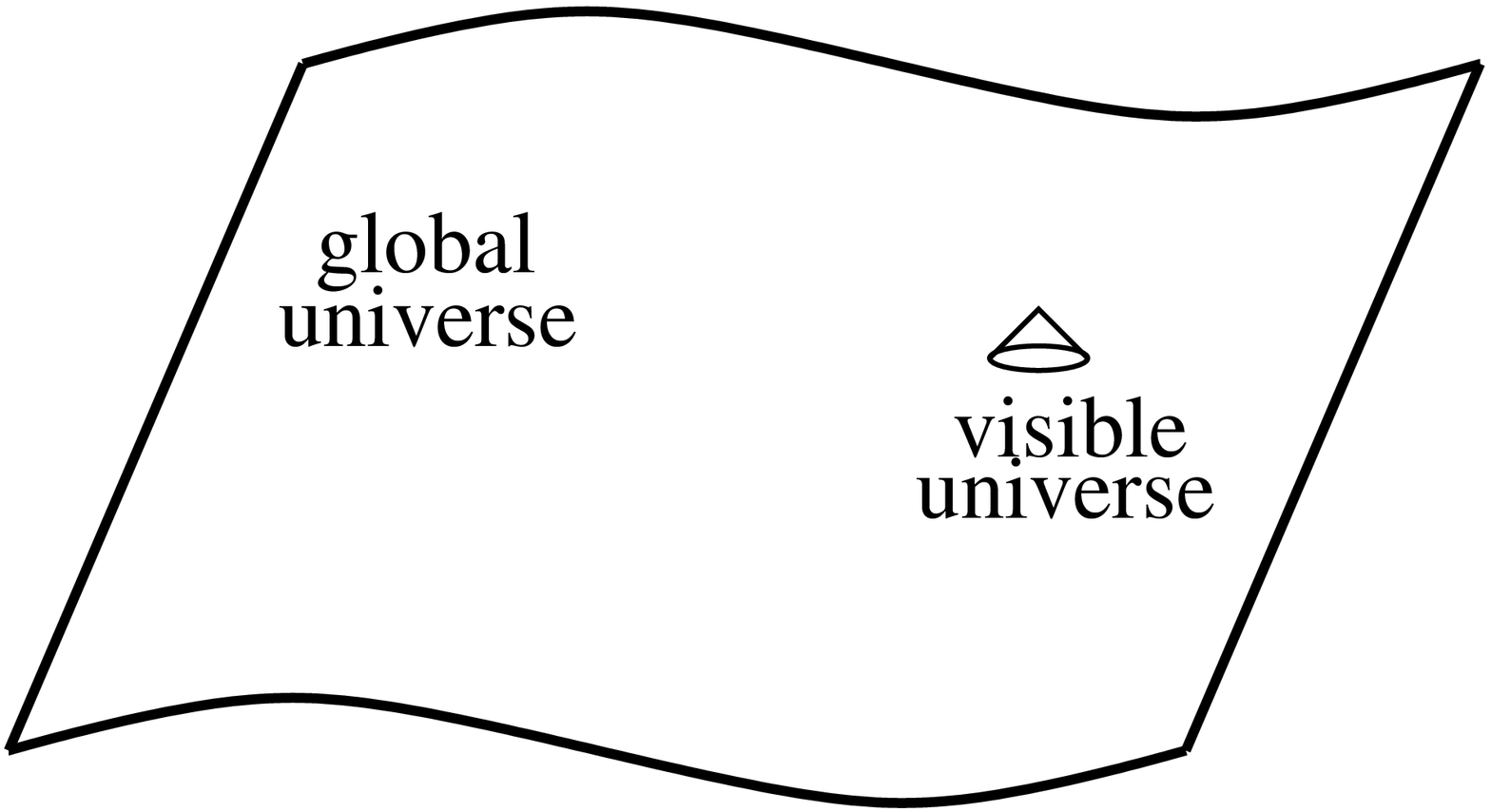,width=8cm}{\label{fig-visible} If the early
  universe underwent inflation, the universe may well be exponentially
  larger than the visible portion (the interior of our past
  lightcone).}

Depending on the energy scale at which inflation occurred, perhaps 60
$e$-foldings suffice to guarantee that $.1\leq\Omega\leq 2$ today.
But it is easy to write down inflationary models with thousands or
millions of $e$-foldings.  In such models, the universe would be spatially
flat not only on the present horizon scale, but on exponentially
larger scales, which will become visible only after an exponentially
longer time than the 13 billion years that have elapsed since the end
of inflation.

The true abundance of such models in the potential landscape we get
from fundamental theory (Sec.~\ref{sec-predict}) is not yet known.
But apparently it is not exceedingly hard to get 60 $e$-foldings, or
else we would have seen curvature long ago.  This suggests that models
with more $e$-foldings are not very rare.  It would seem to require
some tuning for inflation to have lasted just long enough for the
first observable deviations from flatness to occur in the present era.
Thus, most inflationary theorists considered $\Omega=1$ to be a
prediction of inflation.  By the same token, one would expect that the
universe is much larger than the visible universe, perhaps by as much
as $10^{100}$ or $10^{100000}$ (Fig.~\ref{fig-visible}).

\subsection{The cosmological constant problem}
\label{sec-ccp}

When Einstein wrote down the field equation for general relativity, 
\begin{equation}
R_{\mu\nu} - \frac{1}{2} R g_{\mu\nu} + \Lambda g_{\mu\nu} =
8\pi T_{\mu\nu}
\end{equation}
he had a choice: The cosmological constant $\Lambda$ was not fixed by
the structure of the theory.  There was no formal reason to set it to
zero, and in fact, Einstein famously tuned it to yield a static
cosmological solution---his ``greatest blunder''.

The universe has turned out not to be static, and $\Lambda$ was
henceforth assumed to vanish.  This was never particularly satisfying
even from a classical perspective.  The situation is not dissimilar to
a famous problem with Newtonian gravity---that there is no formal
necessity to equate the gravitational charge with inertial mass.

In any case, the simple fact that the universe is large implies that
$|\Lambda|$ is small.  I will show this first for the case of positive
$\Lambda$.  Assume, for the sake of argument, that no matter is
present ($T_{\mu\nu}=0$).  Then the only isotropic solution to
Einstein's equation is de~Sitter space, which exhibits a cosmological
horizon of radius 
\begin{equation}
R_\Lambda= \sqrt{3/\Lambda}~.
\end{equation}
A cosmological horizon is the largest observable distance scale, and
the presence of matter will only decrease the horizon
radius~\cite{Bou00a}.  We see scales that are large in Planck units,
so the cosmological constant must be small in these natural units.

Negative $\Lambda$ causes the universe to recollapse independently of
spatial curvature, on a timescale of order $\Lambda^{-1/2}$.  The
obvious fact that the universe is old compared to the Planck time then
implies that $|\Lambda|$ is small.  

These qualitative conclusions do not require any careful measurements.
Let us plug in some crude numbers that would have been available
already thirty years ago, such as the size of the horizon given in
Eq.~(\ref{eq-h}), or an age of the universe of order $10^{10}$
years.  They imply that
\begin{equation}
  |\Lambda|\lesssim 10^{-122}~.
\label{eq-small}
\end{equation}
Hence $\Lambda$ is very small indeed.

This result makes it tempting to cast scruples aside and simply set
$\Lambda=0$.  But from a modern perspective, to eliminate $\Lambda$ in
the classical Einstein equation is not only arbitrary, but futile.
$\Lambda$ returns through the back door, via quantum contributions to
the stress tensor, $\langle T_{\mu\nu}\rangle$.  It is this effect
that makes the cosmological constant problem so notorious.\footnote{In
  parts, our discussion will follow Refs.~\cite{Wei89,Car00}, where
  more details and references can be found.}

In quantum field theory, the vacuum is highly nontrivial.  Every mode
of every field contributes a zero point energy to the energy density
of the vacuum (Fig.~\ref{fig-loop}a).  The corresponding stress
tensor, by Lorentz invariance, must be proportional to the metric:
\begin{equation}
\langle T_{\mu\nu} \rangle = -\rho_\Lambda g_{\mu\nu} ~.
\label{eq-lrho}
\end{equation}
Though it appears on the right hand side of Einstein's equation,
vacuum energy has the form of a cosmological constant, with
$\Lambda=8\pi\rho_\Lambda$.\footnote{This is why the mystery of the
  smallness of $\rho_\Lambda$ is usually referred to as the
  cosmological constant problem.  But it would be more appropriate to
  call it the vacuum energy problem, since the quantum contributions
  to the vacuum energy are what makes the problem especially hard.}
Its magnitude will depend on the cutoff.
\EPSFIGURE{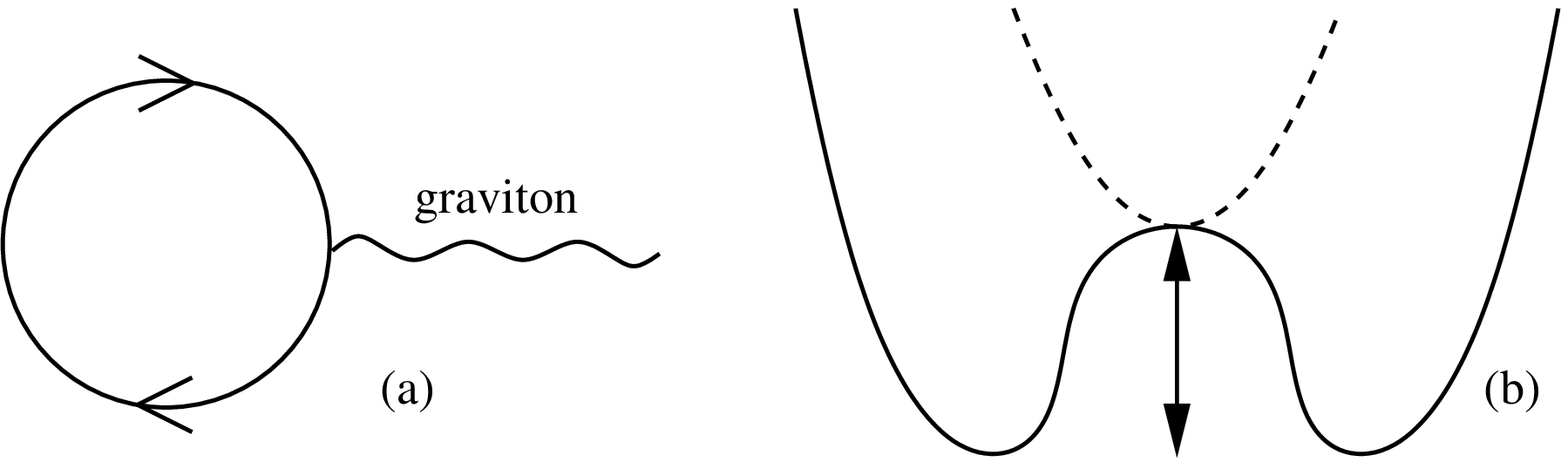,width=.8\textwidth}{\label{fig-loop} Some
  contributions to vacuum energy.  (a) Virtual particle-antiparticle
  pairs (loops) gravitate.  This is mandated by the equivalence
  principle, and has been verified experimentally to a high degree of
  accuracy~\cite{Pol06}.  The vacuum of the standard model abounds with
  such pairs and hence should gravitate enormously. (b) Symmetry
  breaking in the early universe (e.g., the chiral and electroweak
  symmetry) shifts the vacuum energy by amounts dozens of orders of
  magnitude larger than the observed value.}

For example, consider the electron, which is well understood at least
up to energies of order $M=$ 100 GeV.  Dimensional analysis implies
that electron loops up to this cutoff contribute of order $(100$
GeV$)^4$ to the vacuum energy, or $10^{-68}$ in Planck units.  Similar
contributions are expected from other fields.  The real cutoff is
probably of order the supersymmetry breaking scale, giving at least a
TeV$^4\approx 10^{-64}$.  It may be as high as the Planck scale, which
would yield $\Lambda$ of order unity.  Thus, quantum field theory
predicts $\Lambda$ to be some 60 to 120 orders of magnitude larger
than the experimental bound, Eq.~(\ref{eq-small}).

Additional contributions come from the potentials of scalar fields,
such as the potential giving rise to symmetry breaking in the
electroweak theory (Fig.~\ref{fig-loop}b).  The vacuum energy of the
symmetric and the broken phase differ by approximately $(200$
GeV$)^4$.  Any other symmetry breaking mechanisms at higher or lower
energy (such as chiral symmetry breaking of QCD, $(300$ MeV$)^4$) will
also contribute.\footnote{Incidentally, this means that the vacuum
  energy in the early universe was many orders of magnitude larger
  than today.  This follows from well-tested physics and has been
  known for a long time, and it should have made us suspicious of the
  idea that the vacuum energy somehow ``had'' to be exactly zero.  If
  it was ok to have lots of it a few billion years ago, what could be
  fundamentally wrong with having some now?  It also shows that any
  mechanism that would set the vacuum energy to zero in the very early
  universe cannot solve the cosmological constant problem, since
  $|\Lambda|$ would become huge after symmetry breaking.}

I have exhibited various unrelated contributions to the vacuum
energy.  Each is dozens of orders of magnitude larger than the
empirical bound today, Eq.~(\ref{eq-small}).  In particular, the
radiative correction terms from quantum fields are expected to be at
least of order $10^{-64}$.  They can come with different signs, but it
would seem overwhelmingly unlikely for all of them to be carefully
arranged to cancel to such exquisite accuracy ($10^{-122}$) in the
present era.  

This is the cosmological constant problem: why is the vacuum energy
today so small?  It represents an immense crisis in physics: a
discrepancy between theory and experiment, of 60 to 120 orders of
magnitude, in a quantity as basic as the weight of empty space.

\subsection{Strategies and predictions}

Since the 1980s, various strategies for approaching the cosmological
constant problem have been suggested.  They fall into two broad
classes, with each class facing chararacteristic challenges and making
a characteristic prediction.  To give them a fair hearing, let us
assume the cosmological data available in the 1980s: the cosmological
constant is tightly bounded, but has not yet been measured directly.
It might vanish or it might not.

\subsubsection{$\Lambda$ must vanish}
\label{sec-van}

The first approach is to seek a universal symmetry principle that
requires that $\Lambda=0$ in our universe today.  The problem, of
course, is that this challenge has yet to be met.  (Supersymmetry
guarantees that radiative contributions to the cosmological constant
vanish, but in our universe supersymmetry is broken at a scale of at
least a TeV.)  The challenge is not made easier by the fact that one
must allow for a large cosmological constant in the early universe,
when various symmetries were not yet broken.

Assuming these challenges could be met, the first approach does make a
sharp prediction: $\Lambda=0$.

\subsubsection{$\Lambda$ is variable}
\label{sec-var}

The second strategy~\cite{Sak84,Ban85,Wei87} is to posit that the
universe is large---exponentially larger than the presently visible
portion---and that $\Lambda$ varies from place to place, though it can
be constant over very large distances.  As I will explain below,
structure such as galaxies will only form in locations
where~\cite{Wei87}
\begin{equation}
-10^{-123}\lesssim\rho_\Lambda\lesssim  10^{-121}~.
\label{eq-weinberg}
\end{equation}
Since structure is presumably a prerequisite for the existence of
observers, we should then not be surprised to find ourselves in such a
region.

Why is $\Lambda$ related to structure formation?  To form galaxies and
clusters, the tiny density perturbations visible in the cosmic
microwave background radiation had to grow under their own gravity,
until they became non-linear and decoupled from the cosmological
expansion.  This growth is logarithmic during radiation domination,
and linear in the scale factor during matter domination.  Vacuum
energy does not get diluted so it inevitably comes to dominate the
energy density.  As soon as this happens, perturbations cease to grow,
and the only structures that remain gravitationally bound are
overdense regions that have already gone nonlinear.  This means that
there would be no structure in the universe if the cosmological
constant had been large enough to dominate the energy density before
the first galaxies formed~\cite{Wei87}.  This leads to the upper bound
in Eq.~(\ref{eq-weinberg}).  The lower bound comes about because the
universe would have recollapsed into a big crunch too rapidly if the
cosmological constant had been large and negative~\cite{BarTip}.

The problem with the second strategy is twofold:
\begin{enumerate}
\item{It works only in a theory in which $\Lambda$ is a dynamical
    variable whose possible values are sufficiently closely spaced
    that Eq.~(\ref{eq-weinberg}) can be satisfied.}
\item{Assuming generic initial conditions, one would need to find a
    mechanism by which at least one value of $\Lambda$ satisfying
    Eq.~(\ref{eq-weinberg}) can be dynamically attained in a
    sufficiently large region in the universe.}
\end{enumerate}
Supposing that these challenges can be met, one would expect our local
cosmological constant to be fairly typical among the possible values
of $\Lambda$ compatible with structure formation.  In an evenly spaced
spectrum, most values of Lambda satisfying Eq.~(\ref{eq-weinberg})
will be of order $10^{-121}$; for example, only a very small fraction
will be of order $10^{-146}$.

Thus, the ``variable $\Lambda$'' approach predicts~\cite{Wei87} that
the cosmological constant is not much smaller than required by
Eq.~(\ref{eq-weinberg}).  This means that it will be large enough to
be detectable in the present era.  In other words, the ``variable
$\Lambda$'' approach predicts that the vacuum energy should be nonzero
and comparable to the matter density today.

\section{Precision cosmology}
\label{sec-pc}

Beginning with the measurement of anisotropies in the cosmic microwave
background in the 1990s~\cite{COBE}, experimental cosmology has
undergone a remarkable transformation.  The subject has evolved from
order-of-magnitude estimates of a few cosmological parameters to
precise measurements of increasingly complex phenomena, leading to the
emergence of a ``standard model'' of cosmology.  I will not attempt to
review these developments in any detail; see, e.g.,
Refs.~\cite{Teg03,CopSam06,Spe06,SelSlo06,Teg06}.  Instead
I will summarize how several independent types of observations have
helped us evaluate the proposals discussed in the previous section.
This is shown schematically in Fig.~\ref{fig-precision}.
\EPSFIGURE{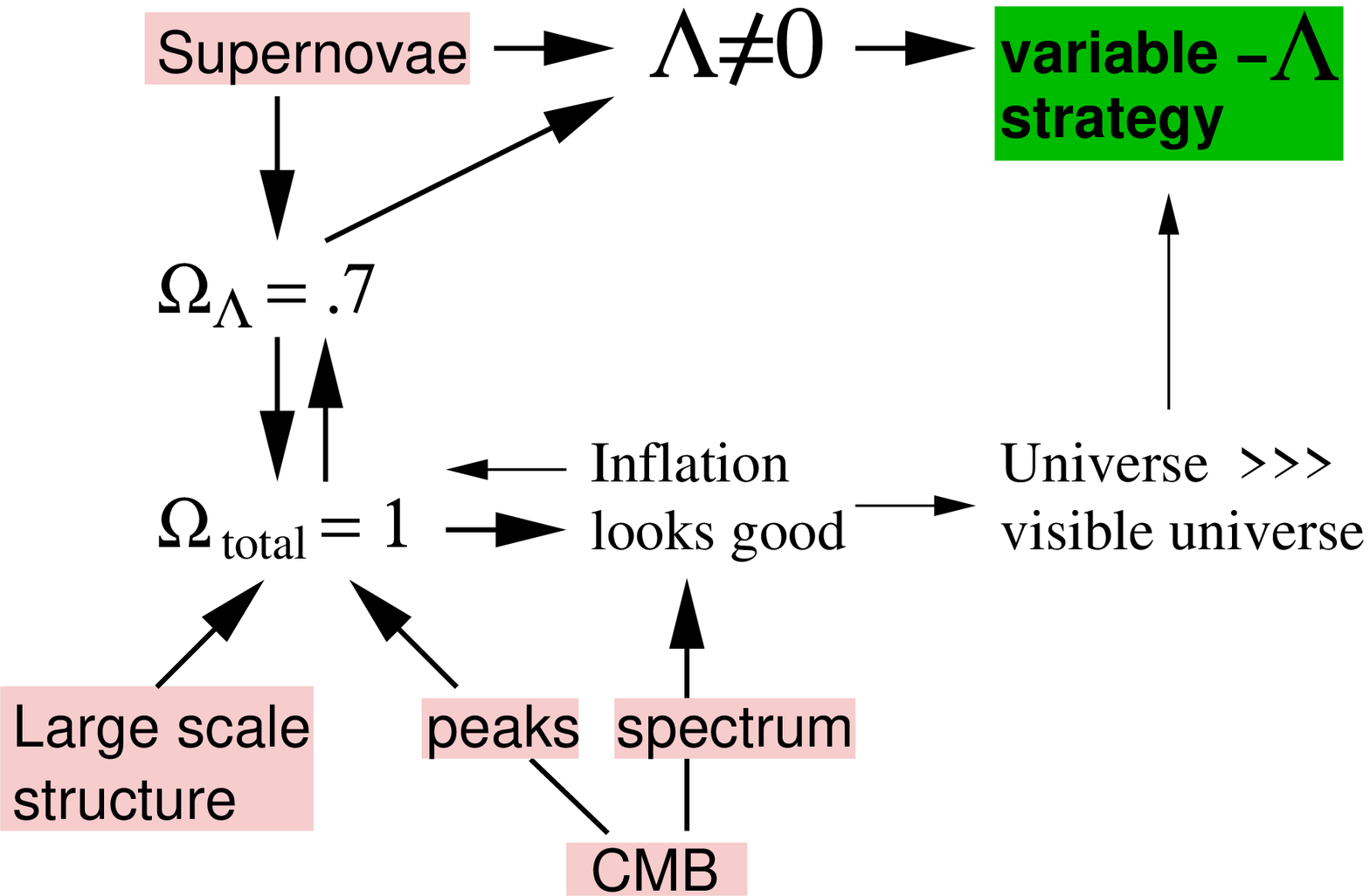,width=.9\textwidth}{\label{fig-precision}
  Recent cosmological precision data (light/red shading) strongly
  support the idea that the cosmological constant is an environmental
  variable that can scan densely spaced values.  The thinner arrows
  indicate that a result merely adds plausibility to another; the
  thicker ones denote the most straightforward implication of a
  result.}

\subsection{Inflation looks good}
\label{sec-infgood}

\begin{enumerate}

\item{Measurements of fluctuations in the cosmic microwave background
    radiation (CMB) strongly support inflation, in two ways:}

\begin{itemize}
\item{The position of the peaks of the perturbation spectrum as a
  function of angular scale imply that the universe is spatially flat
  to excellent precision.  Not only is $\Omega\sim O(1)$, but
  $\Omega=1$, to accuracy of a few percent.  This is the expected
  result if inflation was the correct explanation of the flatness of
  the universe.}
\item{The detailed spectrum of perturbations is nearly scale-invariant
    and Gaussian.  This is natural in inflationary models and rules
    out many other possible seeds for structure formation, such as
    topological defects.}
\end{itemize}

\item{Measurements of the large scale structure (the distribution of
    galaxies and galaxy clusters), by techniques such as weak lensing
    and the Lyman alpha forest, are consistent with $\Omega=1$ and
    have reduced the error bars on this result, supporting inflation.}

\item{Supernova measurements have detected an extra contribution to
    the total energy density in the universe, $\Omega_\Lambda=.7$.
    Meanwhile, the observation of large scale structure has
    corroborated the view that most pressureless matter is dark,
    $\Omega_{\rm matter}=.3$.  This implies independently of the
    previous arguments that $\Omega=1$.}

\end{enumerate}

This evidence directly supports inflation.  Thus, it indirectly lends
credence to the ``variable $\Lambda$'' approach the cosmological
constant problem (Sec.~\ref{sec-var}).  That strategy requires that
the universe be much bigger than what we can presently see of it.  As
I discussed at the end of Sec.~\ref{sec-inflation}, this type of
global picture is natural in inflationary theory.\footnote{In
  Sec.~\ref{sec-predict} I will argue that one should not, in fact,
  attempt to describe all of this global spacetime at once.  Because
  different regions are forever causally disconnected, they correspond
  to different outcomes in a decoherent history.}

\subsection{The cosmological constant is non-zero}
\label{sec-lnonzero}

\begin{enumerate}

\item{Supernova experiments show that the universe began accelerating
  its expansion approximately seven billion years ago.  This indicates
  the presence of vacuum energy with $\rho_\Lambda=1.25\times
  10^{-123}$.  Present data disfavor any time-dependence of this
  component.  Thus, the data strongly support the conclusion that the
  cosmological constant is non-zero.}

\item{The CMB and large scale structure measurements cited in
    Sec.~\ref{sec-infgood} above reinforce this conclusion, since they
    imply that $\Omega=1$.  This value cannot be accounted for by dark
    matter alone.  It implies that at least $70\%$ of the universe
    consists of energy that doesn't clump.  The simplest such
    component is a cosmological constant.}

\item{Indirectly, this conclusion is also supported by the
  measurements of the perturbation spectrum in the CMB cited above.
  They favor inflationary models, and inflation generically predicts
  $\Omega=1$.}

\end{enumerate}

In summary, there is now strong evidence that $\Lambda>0$.  But a
non-zero value of $\Lambda$ in the observed range is precisely what
the ``variable $\Lambda$'' approach to the cosmological constant
problem (Sec.~\ref{sec-var}) predicted.  This is rather fortunate,
since string theory naturally leads to a concrete implementation of
the ``variable $\Lambda$'' strategy, which I will discuss in
Sec.~\ref{sec-bp}.

The data essentially rule out the ``$\Lambda$-must-vanish'' approach
(Sec.~\ref{sec-van}), since $\Lambda$ apparently does not vanish.  But
one could argue that the approach has merely become less appealing,
requiring more epicycles to match observation.  I will now try to
quantify this, before returning to the ``variable $\Lambda$''
strategy.

\subsection{The price of denial}
\label{sec-denial}

The idea that $\Lambda$ is an environmental variable is a perfectly
logical possibility, but it does represent a retreat.  An apt
analogy~\cite{Wei05} is Kepler's hope of explaining the relation
between planetary orbits from first principles.  The hope was dashed
by Newton's theory of gravitation.  Of course, that was no reason to
reject a theory of tremendous explanatory power.  We simply came to
accept that the orbits are the results of historical accidents and
that there are many other solar systems in which different
possibilities are realized.

But let us not be too hasty in abandoning the quest for a unique
prediction of today's value of $\Lambda$.  Instead, let us ask what it
would take to maintain this type of approach in light of the discovery
of non-zero vacuum energy.

We would need to assume that some symmetry or other effect makes
$\Lambda$ vanish, except for a correction of order $10^{-123}$.  This
takes a miracle as the starting point: despite decades of work, no
mechanism has been found that requires $\Lambda=0$ without running
into conflict with known physics~\cite{Wei89,Pol06}.  And supposing it
existed, how would any posited correction evade a mechanism so
powerful as to cancel out many enormous and disparate contributions to
vacuum energy (see Sec.~\ref{sec-ccp})?  Finally, why does this
correction have just the right magnitude so as to be comparable to the
matter density at the present time?

In short, the price of insisting on a unique prediction for the
cosmological constant is that the cosmological constant problem breaks
up into three problems, none of them solved:\footnote{In some
  discussions, the cosmological constant problem is identified with
  these three questions.  But this implicitly assumes that $\Lambda$
  is unique.  Fundamentally, the cosmological constant problem is only
  one question: why is the vacuum energy not huge?  As I explained in
  Sec.~\ref{sec-var}, the ``variable $\Lambda$'' approach predicts
  that $\Lambda$ will be small, but large enough to be already
  noticable in our era.  Thus it avoids the first in our list of
  questions; it answers the third before we have a chance to worry
  about it; and the second question does not arise.  Indeed, at
  present it is senseless to ask why $\Lambda\neq 0$, since we know of
  no reason why $\Lambda$ {\em should\/} vanish.  That it is asked
  anyway betrays only how deeply we had absorbed the prejudice that it
  does.}
\begin{enumerate}
\item{What makes the cosmological constant vanish?}
\item{Why is the cosmological constant not exactly zero?}
\item{Why now?}
\end{enumerate}
The first of these three problems seems by far the hardest; in any
case, it has resisted several decades of attack.  It is tempting to
assume it solved, and to speculate instead about the putative
correction that makes $\Lambda$ nonzero.  But let us be mindful that
any results obtained in this manner will rest on wishful thinking.

Among such approaches, dynamical scalar fields (``quintessence'') take
a prominent role, perhaps because they posit observable deviations
from the equation of state of a cosmological constant.  I confess that
I find this development perplexing.  Dynamical scalars do not match
the data better than a fixed cosmological constant, and they are
theoretically far more baroque.

Scalar fields like to roll off to infinity rapidly, or quickly get
stuck in a local minimum.  For a scalar to mimic vacuum energy and yet
exhibit nontrivial dynamics more than ten billion years after the big
bang, would require an extremely flat (but not exactly flat) potential
over an enormous range.  This necessitates
tunings~\cite{KolLyt98,Car98,Wei00} that include, but go far beyond,
arbitrarily setting the present vacuum energy to a small value.  Yet
further tuning~\cite{Car98} is needed to explain why the long-range
force associated with an almost massless scalar has not been
detected.\footnote{Some authors do confront these latter problems (see
  Refs.~\cite{ChaHal04,Svr06} for recent examples).  Aside from the
  unsolved theoretical question of why $\Lambda$ should vanish at late
  times, such models also receive increasing pressure from
  observation, since dark energy does appear to be at least
  approximately constant.}

Thus, quintessence not only fails to address the very real question of
why $\Lambda$ is small, but, unprovoked by data, burdens us with the
challenge of explaining several additional very small numbers.

Understandably, experimenters demand parametrizations of some spaces
of models that they can hope to constrain~\cite{Alb06}.  But let us
not confuse models (which come cheaper the more complicated we make
them) with explanations.  The danger is that we will forever abuse the
data to constrain ever more baroque models while overlooking the
simplest one~\cite{LidMuk06}.

A cosmological constant is already favored by experiment, and it is
arguably the only model for which we have at least a tentative
fundamental explanation (Sec.~\ref{sec-bp}).  If one finds this
explanation unattractive, it makes sense to seek a different origin of
the simplest model compatible with the data.  What makes no sense is
to write down more complicated models than the data require, while
making no attempt to explain their origin in a credible fundamental
theory.\footnote{Similar remarks apply to the idea that gravity should
  be modified to account for the apparent deviations from
  $\Lambda=0$. This approach also makes sense only to the degree that
  we have any reason to believe that $\Lambda$ should vanish at late
  times, which we don't.  In a modified gravity theory, the quantum
  field theory contributions to the cosmological constant would be
  just as large, unless one violates the equivalence principle, which
  conflicts with other experiments~\cite{Pol06}.}

I am not, of course, proposing that we stop looking experimentally for
any time dependence of dark energy.  The evidence for a nonzero
cosmological constant is surely among the most profound insights ever
gained from experiment.  This alone warrants every effort to confirm
and refine what we know about dark energy.  Perhaps more surprises
await us, complicating the story further.  Meanwhile, I feel that we
theorists would do well to solve the problems we actually have; those
are bad enough.

\section{The discretuum}
\label{sec-bp}

I have argued that experiment favors the ``variable $\Lambda$''
approach to the cosmological constant problem.  I have also
spelled out the main challenges to its implementation.  In this
section, I will present evidence that these challenges are met by
string theory.   Large parts of this section are based on joint work
with J.~Polchinski~\cite{BP}.

\subsection{A continuous spectrum of $\Lambda$?}

The first task is to show that the cosmological constant can take on a
sufficiently dense ``discretuum'' of values.  In string theory, each
line in the spectrum of $\Lambda$ will correspond to a long-lived
metastable vacuum.

Why look for a discretuum and not a continuum of values?  The quick
answer is that we can plausibly realize a discretuum in string theory,
but not a continuum.  In fact, we know of no adjustable parameters on
which the cosmological constant depends in a continuous manner---at
least if our goal at the same time is metastability~\cite{BP}.

But why insist on metastability?  I will give a brief argument that we
have good reasons to do so.  This shows more generally that it would
be difficult to realize the ``variable $\Lambda$'' approach with a
continuous spectrum.

If the continuous parameter is like an integration constant, fixed
once and for all, then it will not allow $\Lambda$ to vary between
large regions in the universe, so it would have to be tuned by hand.
If the parameter can change over time, then the vacuum energy can be
lowered by sliding down the spectrum continuously.  But this is
tantamount to introducing a scalar field potential, and it leads to
versions of problems described in Sec.~\ref{sec-denial}: Why, in ten
billion years, has $\Lambda$ not relaxed to its lowest possible value?
(We cannot assume that this ``ground state'' is the observed value, or
zero, since this would beg the question; radiative corrections would
immediately destroy such a setup.)

It is difficult to see how such a special behavior could be arranged,
other than in a theory with many metastable vacua, but this would get
us back to the discrete case.  Moreover, even with anthropic
constraints there is no reason why $\Lambda$ should change as slowly
as current bounds indicate.  Thus, one would predict a universe with
blatantly time-dependent vacuum energy.  In the discretuum, on the
other hand, the minimum value of the cosmological constant naturally
remains fixed for the lifetime of the metastable vacuum, which can
easily exceed ten billion years.

\subsection{A single four-form field}
\label{sec-single}

To begin, I will present a very simple model of a discretuum.  This
model will not work for two reasons: it cannot be realized in string
theory, and it produces an empty universe~\cite{BT1,BT2}.
Nevertheless, it will be instructive, and it invites a useful analogy
with electromagnetism.

Recall that the Maxwell field, $F_{ab}$, is derived from a potential,
$F_{ab} = \partial_a A_b - \partial_b A_a$.  The potential is sourced
by a point particle through a term $\int e \mathbf A$ in the action,
where the integral is over the worldline of the particle, and e is the
charge.  Technically, $\mathbf F$ is a two-form (a totally
antisymmetric tensor of rank 2), and $\mathbf A$ is a one-form
coupling to a one-dimensional worldvolume (the worldline of the
electron).

The field content of string theory and supergravity is completely
determined by the structure of the theory.  It includes a four-form
field, $F_{abcd}$, which derives from a three-form potential:
\begin{equation}
F_{abcd} = \partial_{[a} A_{bcd]}~,
\end{equation}
where square brackets denote total antisymmetrization.  This potential
naturally couples to a two-dimensional object, a membrane, through a
term $\int q \mathbf A$, where the integral runs over the 2+1
dimensional membrane worldvolume, and $q$ is the membrane charge.

The properties of the four-form field in our 3+1 dimensional world
mirror the behavior of Maxwell theory in a 1+1 dimensional system.
Consider, for example, an electric field between two capacitor plates.
Its field strength is constant both in space and time.  Its magnitude
depends on how many electrons the negative plate contains; thus it
will be an integer multiple of the electron charge: $E = ne$.

Its energy density will be one half of the field strength squared:
\begin{equation}
\rho = \frac{F_{ab}F^{ab}}{2} = \frac{n^2e^2}{2}
\end{equation}
In order to treat this as a system with only one spatial dimension, I
have integrated over the directions transverse to the field lines, so
$\rho$ is energy per unit length.  The pressure is equal to $-\rho$.
The corresponding 1+1 dimensional stress tensor has the form of
Eq.~(\ref{eq-lrho}), so the electromagnetic stress tensor acts like
vacuum energy in 1+1 dimensions.

The same is true for the four-form in our 3+1 dimensional world.
First of all, the equation of motion in the absence of sources is
$\partial_a(\sqrt{-g} F^{abcd}) = 0$, with solution
\begin{equation}
F^{abcd} = c \epsilon^{abcd}~,
\end{equation}
where $\epsilon $ is the unit totally anti-symmetric tensor and $c$ is
an arbitrary constant.  In string theory, there are ``magnetic''
charges (technically, five-branes) dual to the ``electric charges''
(the membranes) sourcing the four-form field.  Then, by an analogue of
Dirac quantization of the electric charge, one can show that $c$ is
quantized in integer multiples of the membrane charge, $q$:
\begin{equation}
c = nq~.
\end{equation}
Note that the actual value of the four-form field is thus quantized,
not only the difference between possible values.

The four-form field strength squares to $F_{abcd}F^{abcd} = 24 c^2$,
and the stress tensor is proportional to the metric, with
\begin{equation}
\rho = \frac{1}{2 \times 4!} F_{abcd}F^{abcd} = \frac{n^2 q^2}{2}
\end{equation}
In summary, the four-form field is non-dynamical, and it contributes
$n^2 q^2/2$ to the vacuum energy.  It is thus indistinguishable from a
contribution to the cosmological constant.

Next, let us include non-perturbative quantum effects.  The electric
field between the plates will be slowly discharged by Schwinger pair
creation of field sources.  This is a process by which a electron and
a positron tunnel out of the vacuum.  Since field lines from the
plates can now end on these particles, the electric field between the
two particles will be lower by one unit [$ne\rightarrow (n-1)e$].  The
particles appear at a separation such that the corresponding decrease
in field energy compensates for their combined rest mass.  They are
then subjected to constant acceleration by the electric field until
they hit the plates.  If the plates are far away, they will move
practically at the speed of light by that time.

For weak fields, this tunneling process is immensely suppressed, with
a rate of order $\exp(-\pi m^2/ne^2)$, where the exponent arises as
the action of a Euclidean-time solution describing the appearance of
the particles.  Thus, a long time passes between creation events.
However, over large enough time scales, the electric field will
decrease by discrete steps of size $e$.  Correspondingly, the 1+1
dimensional ``vacuum energy'', i.e., the energy per unit length in the
electric field, will decrease by a discrete amount $[n^2
  e^2-(n-1)^2e^2]/2 = (n-\frac{1}{2})e^2$.  Note that this step size
depends on the remaining flux.

Precisely analogous nonperturbative effects occur for the four-form
field in 3+1 dimensions.  By an analogue of the Schwinger process,
spherical membranes can spontaneously appear.  (This is the correct
analogue: the two particles above form a zero-sphere, i.e., two
points; the membrane forms a two-sphere.)  Inside this source, the
four-form field strength will be lower by one unit of the membrane
charge [$nq\rightarrow (n-1)q$].  The process conserves energy: the
initial membrane size is such that the membrane mass is balanced
against the decreased energy of the four-form field inside the
membrane.  The membrane quickly grows to convert more space to the
lower energy density, expanding asymptotically at the speed of light.

Membrane creation is a well-understood process described by a
Euclidean instanton, and like Schwinger pair creation, is generically
exponentially slow.  Ultimately, however, it will lead to the
step-by-step decay of the four-form field.  Inside a new membrane, the
vacuum energy will be lower by $(n-\frac{1}{2})q^2$.  

This suggests a mechanism for cancelling off the cosmological
constant.  Let us collect all contributions (see Sec.~\ref{sec-ccp}),
except for the four-form field, in a ``bare'' cosmological constant
$\lambda$.  Generically, $|\lambda|$ should be of order unity (at
least in the absence of supersymmetry), and we will assume without
excessive loss of generality that it is negative.  With $n$ units of
four-form flux turned on, the full cosmological constant will be given
by
\begin{equation}
\Lambda = \lambda+ \frac{1}{2} n^2 q^2
\end{equation}

If $n$ starts out large, the cosmological constant will decay by
repeated membrane creation, until it is close to zero.  The smallest
value of $|\Lambda|$ is attained for the flux $n_{\rm best}$, given by
the nearest integer to $\sqrt{2|\lambda|}/q$. The step size near
$\Lambda=0$ is thus given by $(n_{\rm best}-\frac{1}{2})q^2$.  For
this mechanism to produce a value in the Weinberg window,
Eq.~(\ref{eq-weinberg}), this step size would need to be of order
$10^{-121}$ or smaller.  This requires an extremely small membrane
charge,
\begin{equation}
q< 10^{-121} |\lambda|^{-1/2}
\end{equation}
(the bare cosmological constant $\lambda$ is at best of order one).

This leads to two problems~\cite{BT1,BT2}: the small-charge problem,
and the empty-universe problem.  The membrane charge $q$ is now itself
exceedingly small and thus unnatural.  In particular, despite attempts
in this direction~\cite{FenMar00}, it is not known how to realize such
a small charge in string theory.

Assuming the small-charge problem could be resolved, the mechanism
would lead to a universe very different from ours: it would be devoid
of all matter and radiation.  The point is that small values of
$\Lambda$ are approached very gradually from above.  Thus the universe
is dominated by positive vacuum energy all along, leading to
accelerated expansion.  The exponential suppression of membrane
nucleation events ensures that this expansion goes on long enough to
dilute all matter.  Eventual membrane nucleation decreases the vacuum
energy only by a tiny amount ($10^{-121}$ or less).  At best, this
might reheat the universe to $10^{-30}$, or about
$10^{-2}$eV.\footnote{The actual number is vastly smaller still, since
  most of the energy goes into accelerating the growth of the membrane
  bubble.  This is the reason why the empty-universe problem also
  plagues ``old inflation''~\cite{GutWei83}, even though the jump in
  vacuum energy is considerably larger in that case.}  This falls well
short of the $10$ MeV mark necessary to make contact with standard
cosmology, a theory we trust at least back to nucleosynthesis.

\subsection{Multiple four-form fields}
\label{sec-multiple}

The above problems can be overcome by considering a theory with more
than one species of four-form field.  I will explain why this
situation arises naturally in string theory, but first I will discuss
how multiple four-form fields can produce a dense discretuum without
requiring small charges.

Consider a theory with $J$ four-form fields.  Correspondingly there
will be $J$ types of membrane, with charges $q_1,\ldots,q_J$.  Above
I analyzed the case of a single four-form field; essentially the
conclusions still apply to each field separately.  In particular, each
field strength separately will be constant in 3+1 dimensions,
\begin{equation}
F^{abcd}_{(i)} = n_i q_i \epsilon^{abcd}~,
\end{equation}
and it will contribute like vacuum energy to the stress tensor.

Let us again collect all contributions to vacuum energy, {\em
  except\/} for those from the $J$ four-form fields, in a bare
cosmological constant $\lambda$, which I assume to be negative but
otherwise generic (i.e., of order unity).  Then the total cosmological
constant will be given by
\begin{equation}
\Lambda = \lambda + \frac{1}{2}\sum_{i=1}^J n_i^2 q_i^2~.
\label{eq-lmult}
\end{equation}
This will include a value in the Weinberg window,
Eq.~(\ref{eq-weinberg}), if there exists a set of integers $n_i$ such
that
\begin{equation}
2|\lambda|<\sum n_i^2 q_i^2<2(|\lambda|+ \Delta\Lambda)~,
\end{equation}
where $\Delta\Lambda\approx 10^{-121}$.  

A nice way to visualize this problem is to consider a $J$-dimensional
grid, with axes corresponding to the field strengths $n_i q_i$, as
shown in Fig.~\ref{fig-grid}.  
\EPSFIGURE{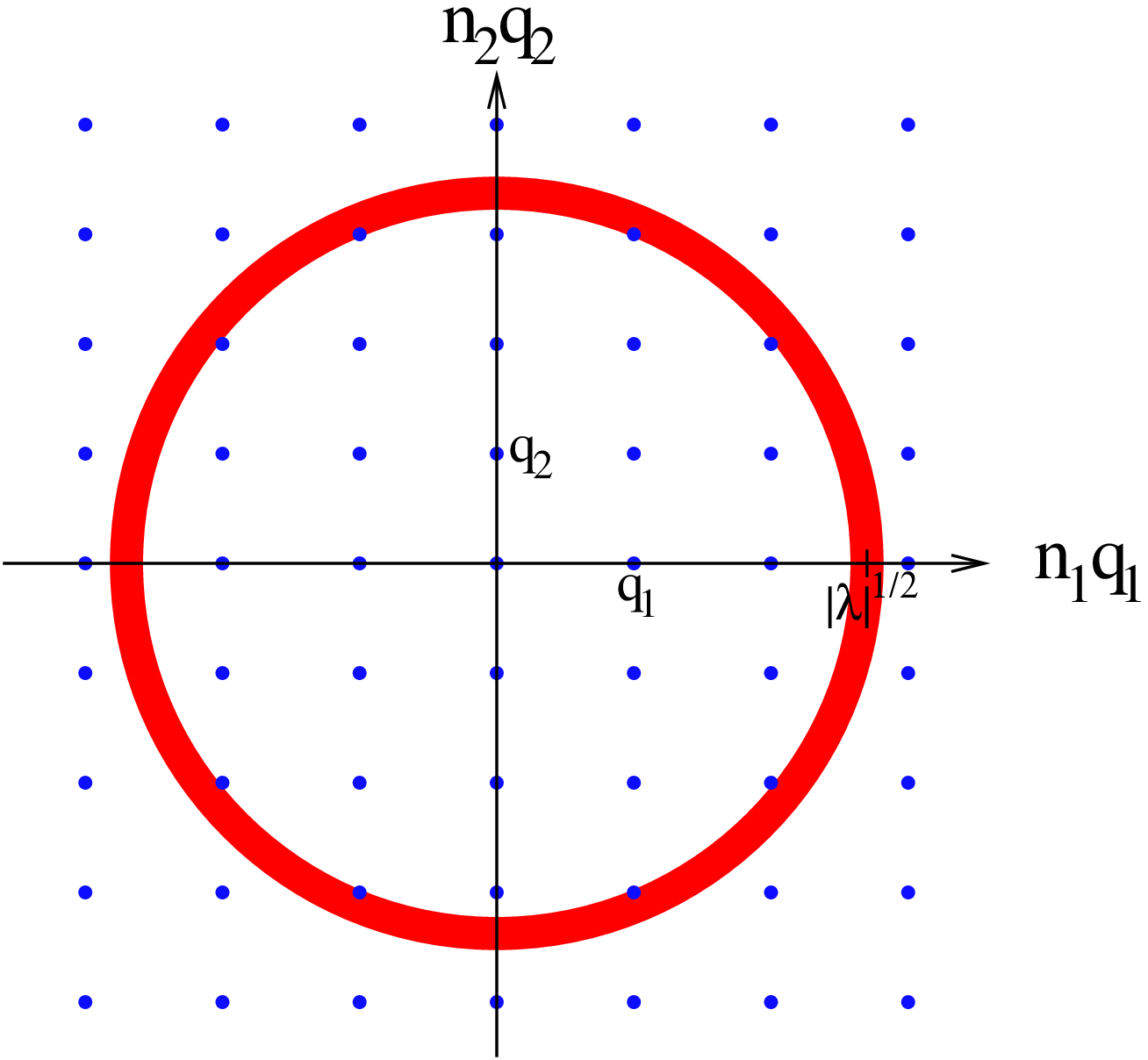,width=.7\textwidth}{\label{fig-grid} Possible
  configurations of the four-form fluxes correspond to discrete points
  in a $J$-dimensional grid.  By Eq.~(\ref{eq-lmult}), vacua that allow
  for structure formation lie within a thin shell of radius
  $\sqrt{2|\lambda|}$ and width $\Delta\Lambda/\sqrt{2|\lambda|}$, where
  $\lambda$ is the bare cosmological constant and $\Delta\Lambda$ is
  the width of the Weinberg window, Eq.~(\ref{eq-weinberg}).}
Every possible configuration of the
four-form fields corresponds to a list of integers $n_i$, and thus to
a discrete grid point.  The Weinberg window can be represented as a
thin shell of radius $\sqrt{2|\lambda|}$ and width
$\Delta\Lambda/\sqrt{2|\lambda|}$.  The shell has volume
\begin{equation}
V_{\rm shell} = \Omega_{J-1} (\sqrt{2|\lambda|})^{J-1}
\frac{\Delta\Lambda}{\sqrt{2|\lambda|}} = \Omega_{J-1}
|2\lambda|^{\frac{J}{2}-1}\Delta\Lambda~,
\end{equation}
where $\Omega_{J-1} = 2\pi^{J/2}/\Gamma(J/2)$ is the area of a unit
$J-1$ dimensional sphere.  The volume of a grid cell is
\begin{equation}
V_{\rm cell}=\prod_{i=1}^J q_i~.
\end{equation}
There will be at least one value of $\Lambda$ in the Weinberg window,
if $V_{\rm cell}<V_{\rm shell}$, i.e., if
\begin{equation}
\frac{\prod_{i=1}^J q_i}{\Omega_{J-1}
  |2\lambda|^{\frac{J}{2}-1}} < |\Delta\Lambda|~.
\label{eq-bp}
\end{equation}

The most important consequence of this formula is that charges no
longer need to be very small.  I will shortly argue that in string
theory one naturally expects $J$ to be in the hundreds.  With $J=100$,
for example, Eq.~(\ref{eq-bp}) can be satisfied with charges $q_i$ of
order $10^{-1.6}$, or $\sqrt{q_i}\approx 1/6$ (the latter has mass
dimension 1 and so seems an appropriate variable for the judging
naturalness of this scenario).  Interestingly, the large expected
value of the bare cosmological constant is actually welcome: it
becomes more difficult to satisfy Eq.~(\ref{eq-bp}) if $|\lambda|\ll
1$.

The origin of the large number of four-form fields lies in the
topological complexity of small extra dimensions.  String theory is
most naturally formulated in 9+1 or 10+1 spacetime dimensions.  For
definiteness I will work with the latter formulation (also known as
M-theory).  If it describes our world, then 7 of the spatial
dimensions must be compactified on a scale that would have eluded our
most careful experiments.  Thus one can write the spacetime manifold as
a direct product:
\begin{equation}
M = M_{3+1}\times X_7~.
\end{equation}
Typically, the compact seven-dimensional manifold $X_7$ will have
considerable topological complexity, in the sense of having large
numbers of non-contractable cycles of various dimensions.

To see what this will mean for the 3+1 dimensional description,
consider a string wrapped around a one-cycle (a ``handle'') in the
extra dimensions.  To a macroscopic observer this will appear as a
point particle, since the handle cannot be resolved.  Now, recall that
M-theory contains five-branes, the magnetic charges dual to membranes.
Like strings on a handle, five-branes can wrap higher-dimensional
cycles within the compact extra dimensions.  A five-brane wrapping a
three-cycle (a kind of non-contractible three-sphere embedded in the
compact manifold) will appear as a two-brane, i.e., a membrane, to the
macroscopic observer.

Six-dimensional manifolds, such as Calabi-Yau geometries, generically
have hundreds of different three-cycles, and adding another dimension
will only increase this number.  The five-brane---one of a small
number of fundamental objects of the theory---can wrap any of these
cycles, giving rise to hundreds of apparently different membrane
species in 3+1 dimensions, and thus, to $J\sim O(100)$ four-form
fields, as required.

The charge $q_i$ is determined by the five-brane charge (which is set
by the theory to be of order unity), the volume of $X_7$, and the
volume of the $i$-th three-cycle.  The latter factors can lead to
charges that are slightly smaller than 1, which is all that is
required.  Note also that the volumes of the three-cycles will
generically differ from each other, so one would expect the $q_i$ to
be mutually incommensurate.  This is important to avoid huge
degeneracies in Eq.~(\ref{eq-lmult}).

Each of the flux configurations ($n_1,\ldots,n_J$) corresponds to a
metastable vacuum.  Fluxes can only change if a membrane is
spontaneously created.  As discussed in Sec.~\ref{sec-single}, this
Schwinger-like process is generically exponentially suppressed,
leading to extremely long lifetimes.  Thus, multiple four-forms
naturally give a dense discretuum of metastable vacua.

The model I have presented is an oversimplification.  When it was
first proposed, it was not yet understood how to stabilize the compact
manifold against deformations (technically, how to fix all moduli
fields including the dilaton).  This is clearly necessary in any case
if string theory is to describe our world.  But one would expect that
in a realistic compactification, the fluxes wrapped on cycles should
deform the compact manifold, much like a rubber band wrapping a
doughnut-shaped balloon.  Yet, I have pretended that $X_7$ stays
exactly the same independently of the fluxes $n_i$.

Therefore, Eq.~(\ref{eq-lmult}) will not be correct in a more
realistic model.  The charges $q_i$, and indeed the bare cosmological
constant $|\lambda|$, will themselves depend on the integers $n_i$.
Thus the cosmological constant may vary quite unpredicably.  But the
crucial point remains unchanged: the number of vacua, $N$, can be
extremely large, and the discretuum should have a typical spacing
$\Delta\Lambda\approx 1/N$.  For example, if there are $500$
three-cycles and each can support up to 9 units of flux, there will be
of order $N=10^{500}$ metastable configurations.  If their vacuum
energy is effectively a random variable with at most the Planck value
($|\Lambda|\lesssim 1$), then there will be $10^{380}$ vacua in the
Weinberg window, Eq.~(\ref{eq-weinberg}).

In the meantime, there has been significant progress with stabilizing
the compact geometry (e.g., Refs.~\cite{DasRaj99,GidKac01}; see
Refs.~\cite{Sil04,Gra05,DouKac06} for reviews.).  In particular,
Kachru, Kallosh, Linde, and Trivedi~\cite{KKLT} have shown that
metastable de~Sitter vacua can be realized in string theory while
fixing all moduli.\footnote{Constructions in non-critical string
  theory (i.e., string theory with more than ten spacetime dimensions)
  were proposed earlier~\cite{Sil01,MalSil02}.}  Their construction
supports the above argument that the number of flux vacua can be
extremely large.  More sophisticated counting methods~\cite{DenDou04b}
bear out the quantitative estimates obtained from the simple model I
have presented.

I will close with two remarks.  The need for extra dimensions could be
regarded as an unpleasant aspect of string theory, since it forces us
to worry about why and how they are hidden.  Ironically, they are
precisely what has allowed string theory to address the cosmological
constant problem and pass its first observational test.

One often hears that there are now $10^{500}$ ``string theories'',
suggesting a loss of fundamental simplicity and uniqueness.  This is
like saying that there are myriads of standard models because there
are many ways to make a lump of iron.  From five standard model
particles, one can construct countless metastable configurations of
atoms, molecules, and condensed matter objects.  Similarly, the large
number of vacua in string theory arises by combining a small set of
fundamental ingredients in different ways, {\em in the extra
  dimensions}.  From this perspective, numbers like $10^{500}$ should
not surprise us.

\subsection{Our way home}
\label{sec-home}

I have argued that string theory contains such a dense spectrum of
metastable vacua that many of them will satisfy the Weinberg
inequality, Eq.~(\ref{eq-weinberg}).  But still, they represent only a
very small fraction of the total number of vacua.  Hence, there is no
particular reason to assume that the universe would have started out
in one of the relatively rare vacua with small late-time cosmological
constant.  Such an assumption would be especially problematic since
the late-time value of the cosmological constant is initially far from
apparent.  In our own vacuum, for example, the cosmological constant
is now small but was enormously larger at early times, before
inflation ended and various symmetries were broken.

Fortunately, it is unnecessary to assume that the universe starts out
in a Weinberg vacuum.  I will now show that starting from generic
initial conditions, the universe will grow arbitrarily large.  Over
time, it will come to contain enormous regions (``bubbles'' or
``pockets'') corresponding to each metastable vacuum
(Fig.~\ref{fig-global}).  In particular, the Weinberg vacua will be
realized somewhere in this ``multiverse''.  It will be seen that these
vacua can be efficiently reheated, so the empty-universe problem of
Sec.~\ref{sec-single} will not arise.
\EPSFIGURE{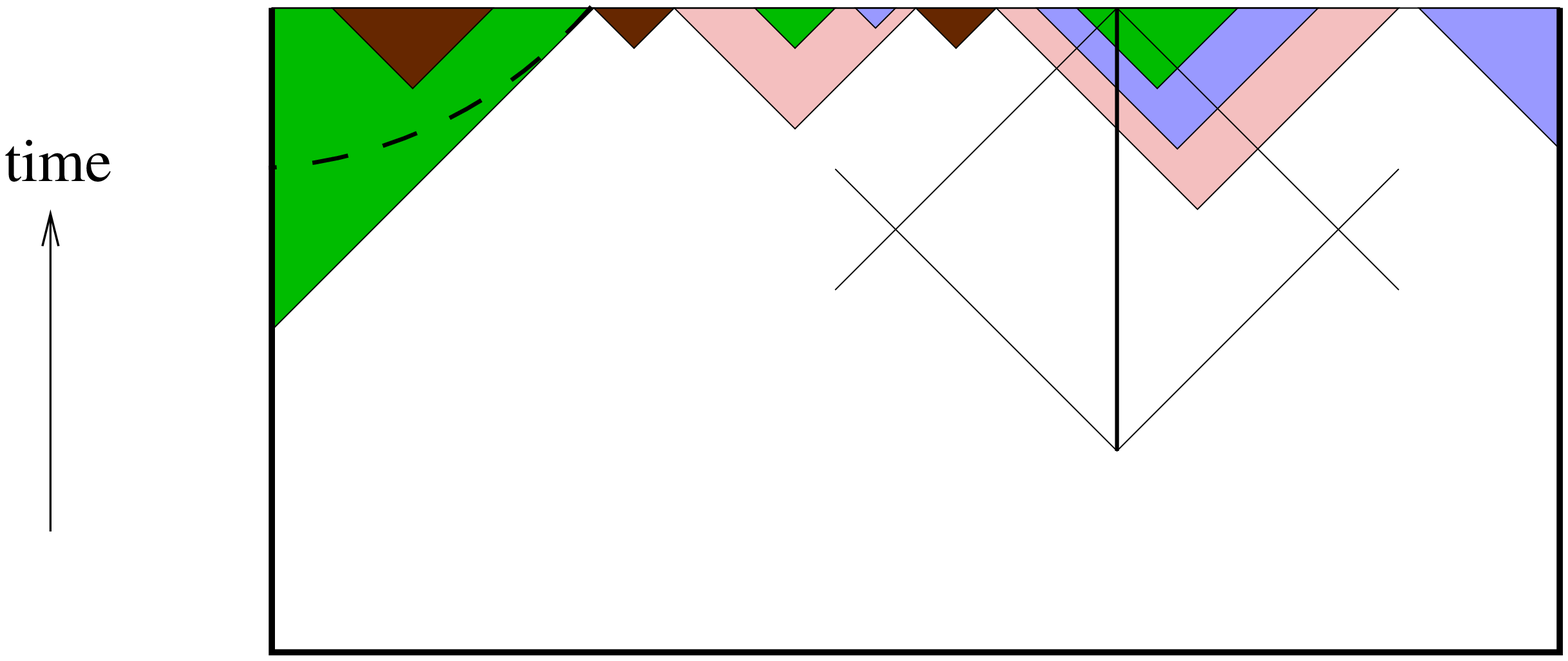,width=.8\textwidth}{\label{fig-global}
  Bird's eye view of the universe.  There are regions corresponding to
  every vacuum in the landscape (shown in different colors).  Each
  region is an infinite, spatially open universe; the dashed line
  shows an example of an instant of time.  The black diamond is an
  example of a spacetime region that is causally accessible to a
  single observer (see Sec.~\ref{sec-predict}).}

By Eq.~(\ref{eq-lmult}), all but a finite number of metastable vacua
will have $\Lambda>0$.  Let us assume that the universe begins in one
of these vacua.  Of course, this means that typically the cosmological
constant will be large initially.  Since $\Lambda>0$, the universe
will be well described by de~Sitter space.  It can be thought of as a
homogeneous, isotropic universe expanding exponentially on a
characteristic time scale $\Lambda^{-1/2}$.

Every once in a long while (this time scale being set by the action of
a membrane instanton, and thus typically much larger than
$\Lambda^{-1/2}$), a membrane will spontaneously appear and the
cosmological constant will jump by $(n_i-\frac{1}{2}) q_i^2$.  But
this does not affect the whole universe.  $\Lambda$ will have changed
only inside the membrane bubble.  This region grows arbitrarily large
as the membrane expands at the speed of light.  

But crucially, this does {\em not}\/ imply that the whole universe is
converted into the new vacuum~\cite{ColDel80}.  This technical result
can be understood intuitively.  The ambient, old vacuum is still, in a
sense, expanding exponentially fast.  The new bubble eats up the old
vacuum as fast as possible, at nearly the speed of light.  But this is
not fast enough to compete with the background expansion.

More and more membranes, of up to $J$ different types, will nucleate
in different places in the rapidly expanding old vacuum.  Yet, there
will always be some of the old vacuum left.  One can show that the
bubbles do not ``percolate'', i.e., they will never eat up all of
space~\cite{GutWei83}.  Thus different fluxes can change, and
different directions in the $J$-dimensional flux space are explored.

Inside the new bubbles, the game continues.  As long as $\Lambda$ is
still positive, there is room for everyone, because the background
expands exponentially fast.  In this way, all the points in the flux
grid $(n_1,\ldots,n_J)$, are realized as actual regions in physical
space.  The cascade comes to an end wherever a bubble is formed with
$\Lambda<0$, but this affects only the interior of that particular
bubble (it will undergo a big crunch).  Globally, the cascade
continues endlessly.

Perhaps surprisingly, each bubble interior is an open FRW universe in
its own right, and thus infinite in spatial extent.\footnote{In an
  open universe, spatial hypersurfaces of constant energy density are
  three-dimensional hyperboloids.  This shape is dictated by the
  symmetries of the instanton describing the membrane nucleation.  It
  is closely related to the hyperbolic shape of the spacetime paths of
  accelerating particles, like the electron-positron pair studied
  above.} Yet, each bubble is embedded in a bigger universe (sometimes
called ``multiverse'' or ``megaverse''), which is extremely
inhomogeneous on the largest scales.

An important difference to the model with only one four-form is that
the vacua will not be populated in the order of their vacuum energy.
Two neighboring vacua in flux space (i.e., neighbors in the
``landscape''), will differ hugely in cosmological constant.  That is,
they differ by one unit of flux, and the charges $q_i$ are not much
smaller than one, so by Eq.~(\ref{eq-lmult}) this translates into an
enormous difference in cosmological constant.  Conversely, vacua with
very similar values of the cosmological constant will be well
separated in the flux grid (i.e., far apart in the landscape).

This feature is crucial for solving the empty universe problem.  When
our vacuum was produced in the interior of a new membrane, the
cosmological constant may have decreased by as much as $1/100$ of the
Planck density.  Hence, the temperature before the jump was enormous
(in this example, the Gibbons-Hawking temperature of the corresponding
de~Sitter universe would have been of order $1/10$ of the Planck
temperature), and only extremely massive fields will have relaxed to
their minima.  Most fields will be thermally distributed and can only
begin to approach equilibrium after the jump decreases the vacuum
energy to near zero.

Thus, the final jump takes on the role analogous to the big bang in
standard cosmology.  The ``universe'' (really, just our particular
bubble) starts out hot and dense.  If the effective theory in the
bubble contains scalar fields with suitable potentials, there will be
a period of slow-roll inflation as their vacuum energy slowly relaxes.
(This was apparently the case in our vacuum.)  At the end of this
slow-roll inflation process, the universe reheats.

To a (purely hypothetical) observer in the primordial era of a given
bubble, it would be far from obvious what the late-time cosmological
constant will be, since this depends on future symmetry breakings and
the relaxation of scalar field potential energy.  The small late-time
values in some bubbles are the result of purely accidental
cancellations---which are bound to happen in some vacua if there are
$10^{500}$ vacua in total.

To a hypothetical primordial observer in our own bubble, the evolution
of vacuum energy would seem like a sequence of bizzare coincidences.
I assume here that the observer is sufficiently intelligent to know
that quantum field theory predicts a cosmological constant of order
one.  In the primordial era, the energy density in radiation is large,
and it could mask even a fairly large cosmological constant.  But as
the universe cools off, a cosmological constant exceeding the ever
decreasing energy density in matter and radiation would become
immediately apparent.  Thus, the discrepancy between theory and
observation grows larger and larger.

Much to his surprise, our observer would find the vacuum energy in the
minimum of the inflationary potential to be much smaller than during
inflation---in fact, it cannot be distinguished from zero.  (This
allows the universe to reheat, without immediately inflating all
matter away, but why would our observer care?)  During electroweak
symmetry breaking, at time $10^{-12}$ sec, the vacuum energy density
shifts by $(200$ GeV$)^4$.  Our observer computes this and is thus led
to expect that soon afterwards, when the radiation energy drops below
$(200$ GeV$)^4$, the dynamical effects of a cosmological constant will
finally become apparent.  It does not, so the observer is forced to
conclude that the shift must have cancelled against another, equally
large contribution that he had not noticed earlier since radiation was
too dense.  In fact, the cancellation is so exquisite that vacuum
energy remains dynamically irrelevant at the much later time $1$ sec.
(This allows nucleosynthesis to proceed.)  After hundreds of millions
of years, at vastly lower energy density, still no vacuum energy is
apparent (allowing for the formation of galaxies to proceed
undisturbed).  Only after billions of years (after structure has
formed), does vacuum energy resurface and begin to dominate over the
ever more dilute matter energy density.

If such hypothetical observers existed, this sequence really {\em
  would\/} be bizzare and unexpected.  There are far more vacua with
similar primordial evolution but without the anomalously small
late-time cosmological constant.  All the corresponding bubbles would
presumably harbor similar primordial observers.  Then the vast
majority of observers would {\em not\/} see a sequence of ``miracles''
leading to a late-time cosmological constant as small as $10^{-121}$.

But it appears that no such hypothetical primordial observers exist.
Observers will arise only after some structure has formed.  This
happens only in the ``bizzare'', rare vacua in which accidental
cancellations produce a late-time cosmological constant of order
$10^{-121}$ or less.  Any larger, and vacuum energy would disrupt
galaxy formation.  We should not be surprised, therefore, to find
ourselves in such a bubble.

\section{The landscape and predictivity}
\label{sec-predict}

\subsection{A new challenge}

A good explanation will do more than solve a problem.  It should offer
us a new way of thinking, and in doing so, raise new, interesting
problems.  In fact, the picture I have outlined does present a
tremendous challenge: how does one make predictions in the landscape?

Let us suppose that there are $10^{500}$ metastable vacua.  Among
them, everything varies: forces, coupling strengths, masses, field
content, gauge groups, and other aspects of the low energy
Lagrangian.  Are the ``constants'' of nature we measure
constrained by nothing but the fact of our existence?  This would be a
bleak prospect indeed.

In order to look at the problem dispassionately, it helps to take
recourse once more to the analogy with complex, many-particle systems
developed near the end of the previous section.  A vast number of
phenomena arise from a few particles in the standard model: the world
is a rich, complex place.  But this does not imply that anything goes.
There are only a finite number of elements, and a random combination
of atoms is unlikely to form a stable molecule.  Even quantities such
as material properties ultimately derive from standard model
parameters and cannot be arbitrarily dialed.  

Similarly, one would expect that there are low-energy Lagrangians that
simply cannot arise from string theory with its limited set of
ingredients, no matter how complicated the manner in which they are
combined~\cite{Vaf05,ArkMot06,OogVaf06}.

Moreover, the great complexity of a system need not be an obstacle to
its effective description.  Imagine we had never heard of
thermodynamics and were told to describe the behavior of all the air
molecules in a room.  Or suppose we were ignorant of condensed matter
physics, and were charged with deriving the properties of metals from
the standard model.  Would we not worry, for a moment, that these
tasks are too complex to be tractable?  Of course, we know well that
such problems yield to the laws of large numbers.  The predictive
power of statistical or effective theories is completely deterministic
in practice: not in ten billion years will the air ever collect in one
corner of the room.  This is not to say that finding such descriptions
is trivial, only that it is possible.

Similarly, there is every reason to hope that a set of $10^{500}$
vacua will yield to statistical reasoning, allowing us to extract
predictions.  Yet we must not presume this task simple or even
straightforward.  We are just beginning, so the present scarcity of
predictions is hardly proof of their impossibility.

The problem can be divided into three separate tasks:
\begin{enumerate}
\item{Statistical properties of the string theory landscape}
\item{Selection effects from cosmological dynamics}
\item{Anthropic selection effects}
\end{enumerate}
The first of these has been tackled by a number of authors; see, e.g.,
Refs.~\cite{Dou03,DenDou04b,GmeBlu05,DouTay06}, or Ref.~\cite{Kum06}
for a review.  The question is, what is the relative abundance of
stable or metastable vacua with specified low-energy properties.  Our
understanding of metastable vacua is still rather qualitative, so most
investigations focus on supersymmetric vacua instead, which are under
far better control.  Clearly, it would be desirable to extend our
samples; this will likely require significant progress in
understanding vacua without supersymmetry.  Meanwhile it will be
interesting to understand the extent to which current samples are
representative of more realistic vacua, especially since one is
usually working in a particular corner of moduli space.

This remains a very active area of research, and I will not attempt a
more detailed review.  Next, I will discuss a recent approach to the
second and third task.

\subsection{Probabilities in eternal inflation}

It is not enough to calculate the probability that a random metastable
vacuum picked from the theory landscape has a given property.
Cosmological dynamics is interposed between the theory landscape and
the actual realization of vacua as large regions in the universe.
This dynamical process may preferentially produce some vacua and
suppress others.  This is the second question listed above: What is
the relative abundance of different vacua {\em in the physical
  universe\/}?

Computational difficulties aside, this question turns out to be hard
to answer even in principle, because of a scourge of infinities.  The
global structure of the universe arising from the string landscape is
extremely complicated (see Sec.~\ref{sec-home}).  Each vacuum $i$ is
realized infinitely many times as a bubble embedded in the global
spacetime.  Moreover, every bubble is an open universe and thus of
infinite spatial extent.

The most straightforward way of regulating the infinities is to
consider the universe at finite time before taking a limit.  There is
an ambiguity in whether one should compare the volumes, or simply the
number of each type of bubble on this time slice (or some intermediate
quantity).  Worse, results depend strongly on the choice of time
variable~\cite{LinLin94,GarLin94}, and no preferred time-slicing is
available in the highly inhomogeneous global spacetime.

A number of slicing-invariant probability measures have been proposed;
see, e.g.~\cite{GarVil01,GarSch05,EasLim05} for recent work.  Yet,
slicing invariance is far from a strong enough criterion for
determining a unique measure; for example, any function of an
invariant measure will again be invariant.

In addition to these severe ambiguities, known slicing-invariant
proposals appear to lead to predictions that disagree with
observation~\cite{FelHal05,GarVil05,Pag06,BouFre06b}.  The first
problem arises in proposals where the probability carried by a vacuum
is proportional to the factor by which inflation increases the volume.
(This refers to the ordinary slow-roll inflation of
Sec.~\ref{sec-inflation}, not the false-vacuum driven eternal
inflation of Sec.~\ref{sec-home}.)  This factor is exponential in the
duration of inflation.  In Ref.~\cite{FelHal05} it was argued that
generically, both the number of $e$-foldings and the density
perturbations produced will depend monotonically on parameters of the
inflationary model.  Thus, the great weight carried by long periods of
inflation should push the density contrast $\delta\rho/\rho$ towards 0
or 1.  One can argue that life would be impossible in a universe with
$\delta\rho/\rho$ too small or too large~\cite{TegRee97}.  But
anthropic arguments cannot resolve the paradox.  The exponential
preference for extreme values means that we should live dangerously, a
lucky fluctuation in an inhospitable universe.  Instead, the density
contrast in our universe appears to be comfortably within the
anthropic window.

A more severe problem arises, e.g., in the proposal by Garriga et
al.~\cite{GarSch05}: One can show that the overwhelming majority of
observers are not like us but arise from random
fluctuations~\cite{Pag06}.  Assuming that we are typical observers (as
we must if we want to make any predictions), this conflicts with
observation.  It could be avoided if all vacua that can harbor
observers decay on a timescale not much longer than $\Lambda^{-1/2}$.
But this is extremely implausible in the string
landscape~\cite{BouFre06b}.

Recently, a local (or ``causal'', or ``holographic'') approach has
been developed which avoids the ambiguities and resolves the paradoxes
described above~\cite{Bou06,BouFre06,BouFre06b,BouHar06}.  Its
original motivation, however, comes from the study of black hole
evaporation, which appears to be a unitary
process~\cite{StrVaf96,Mal97}.  A different kind of paradox arose in
this context: The initial quantum state is duplicated, appearing at
the same instant of time both in the Hawking radiation and inside the
black hole.  However, causality prevents any observer from seeing both
copies.  Thus, the black hole paradox is resolved if we give up on
trying to describe the spacetime globally~\cite{SusTho93,Pre92}.
Indeed, all that is needed is a theory that can describe the
experience of any observer (as opposed to a theory describing
correlations between points remaining forever out of causal contact,
making predictions which cannot be verified even in principle).  But
if the global point of view must be rejected in the context of black
holes, why should it be retained in cosmology?

From a local point of view, eternal inflation looks quite
different~\cite{BouFre06}.  Let us attempt to describe only a single
(though arbitrary) causally connected region.  This can be defined as
a ``causal diamond'': the overlap between the causal future and the
causal past of a worldline~\cite{Bou00a}.  As seen in
Fig.~\ref{fig-global}, this restriction eliminates most of the global
spacetime.  In particular, eternal inflation is no longer eternal.

Consider a geodesic worldline, starting in some initial vacuum $o$
with large positive cosmological constant.  (Really, I am considering
an ensemble of worldlines and regions causally connected to them, in
the sense usually adopted to give meaning to probabilities in quantum
mechanics: identical copies of a system.  I am {\em not\/} demanding
that the members of this ensemble coordinate their evolution so as to
fit together and form a well-defined global spacetime.)  Since the
probability to do so is nonzero, the worldline eventually enters a
vacuum of zero or negative cosmological constant, from which it will
decay no further.\footnote{If $\Lambda$ vanishes exactly then the
  vacuum is presumably supersymmetric and stable.  If $\Lambda<0$ the
  open universe collapses in a big crunch after a time of order
  $\Lambda^{-1/2}$, which is likely to be faster than any further
  decay channels.}  But which vacua the worldline passes through, on
its way to a ``terminal'' vacuum, is a matter of probability.  

The probability for the worldline to enter vacuum $i$, $p_i$, is
proportional to the expected number of times it will enter vacuum $i$.
This can be computed straightforwardly, and unambiguously, from the
matrix of transition rates between vacua~\cite{Bou06}.  

The probabilities $p_i$ depend on the initial probability distribution
for the vacuum in which the worldline starts out, as one would expect
in most dynamical systems.  Inflation does not remove the need for a
theory of initial conditions.  I will not address this question here,
except to say that I find it plausible that the universe began in a
vacuum with large cosmological constant, and was equally likely to
start in any such vacuum.  The vast majority of vacua will have large
cosmological constant, so this is not a strong assumption.

The resulting probability measure is predictive.  In the semiclassical
regime, decays tend to be exponentially suppressed, so that one decay
channel typically dominates completely in any given vacuum.  One would
expect that a number of decays have to happen before the worldline
enters a vacuum on the Weinberg shell, and that the fast decays happen
first.  For example, in a model of the type described in
Sec.~\ref{sec-multiple}, the production of a membrane of type $i$ is
less suppressed if the background has more than one unit of the
corresponding flux ($n_i>1$), or if the charge $q_i$ associated with
the membrane is relatively small.  One thus predicts that the number
of units of flux should be $0$ or $1$ for most fluxes in our vacuum,
and that we are unlikely to find fluxes associated to small charges
turned on~\cite{BouYan06}.

The paradox of Ref.~\cite{FelHal05} is resolved because the size of
the causal diamond is cut off by the cosmological constant.  It will
never become larger than the horizon in a given vacuum, no matter how
much slow-roll inflation occurs after the corresponding bubble is
formed.  Thus, exponentially large expansion factors do not enter.
This does not mean that the measure is insensitive to the important
question of whether inflation occurs.  However, that issue arises only
if we ask about the suitability of vacua for observers.  I will turn
to this question next.

\subsection{Beyond the anthropic principle}

Most vacua will not contain observers.  This statement is not
particularly controversial: for example, most vacua will have a
cosmological constant of order unity, and hence will not give rise to
causally connected regions much larger than a Planck length.  Entropy
bounds~\cite{CEB1,CEB2} imply that such regions contain at most a few
degrees of freedom, and only a few bits of information.  This rules
out complex structures.

Therefore, the probability for a worldline to enter a given vacuum,
$p_i$, is not the same thing as the probability for that vacuum to be
observed, $\pi_i$.  Let us define a weight $w_i$ that measures (in a
sense to be quantified below) the chance that the vacuum $i$ contains
observers.  Then
\begin{equation}
\pi_i = \frac{p_i w_i}{\sum p_j w_j}~.
\end{equation}

Estimating the weights $w_i$ is awkward for a number of reasons.  The
biggest difficulty is to define what we mean by an ``observer''.  And
given a definition, it can still be extremely hard to estimate whether
observers will form in a given vacuum.  What we can do reasonably well
is to consider hypothetical, small changes of one or two of the
parameters describing our own vacuum, and compute their effect on the
formation of life like ours.  But this is of little use for estimating
the weights $w_i$ of other vacua in the landscape, since they
generically have radically different low-energy physics.  Some
correlations may appear quite robust, such as Weinberg's assertion
that some kind of structure formation is a prerequisite for observers.
But others seem hopelessly specific.  For example, can we seriously
expect that life requires carbon?  What would this statement even mean
in a low-energy theory with a different standard model gauge group?

In the global approach, an additional difficulty arises: Strictly,
$w_i$ is either $0$ or $1$.  Either there are observers in vacuum $i$,
or there are not.  Intuitively, this seems too crude; there should be
a more nuanced sense in which some vacua can be more or less
hospitable to life.  But how would we tell whether a vacuum contains
more observers than another?  Each bubble is an infinite homogeneous
open universe.  At all times, the spatial volume is strictly infinite.
So if observers can form at all, there will be an infinite number of
them.  (A method for dealing with this problem within the global
apprach has been suggested in Ref.~\cite{GarVil01}.)

In the local approach, this problem does not arise.  The causal
diamond will be at most of linear size $|\Lambda_i|^{-1/2}$, where
$\Lambda_i$ is the cosmological constant in vacuum $i$.  (I will
ignore vacua with vanishing cosmological constant, since they would
have to be exactly supersymmetric, ruling them out as hosts of complex
structures.)  Thus, the causally connected region is automatically
finite, providing a natural cutoff.

The local approach can also help overcome the problem of the excessive
specificity of anthropic considerations~\cite{Bou06}.  The key idea is
that observers, whatever they may consist of, need to be able to
increase the entropy.  It is implausible that complex systems like
observers will still operate when everything has thermalized and all
free energy has been used up.  Everything interesting happens while
the universe returns to equilibrium after the phase transition
associated with the formation of a new bubble.\footnote{This is the
  reason why I defined the $p_i$ to be the probability for the
  worldline to {\em enter\/} vacuum $i$, rather than the expected
  amount of time the worldline will spend in vacuum $i$.  The latter
  will typically be exponentially greater than the thermalization time
  scale and hence is of no relevance.}

Let us assume that every binary operation will increase the entropy by
at least an amount of order unity~\cite{KraSta00}.  On average, one
would expect the number of observers to be related to the total amount
by which entropy increases in a given vacuum.  Of course, in the
global viewpoint this statement would be nonsense: if the entropy
increases at all, it will increase by an infinite amount over the
infinite open space.  In the local viewpoint, the entropy increase is
not only finite but can be very sharply defined in terms of the causal
diamonds themselves.

The entropy increase is the difference between the entropy entering
the diamond through the bottom cone, $S_{\rm in}$, and the entropy
leaving through the top cone, $S_{\rm out}$, as shown in
Fig.~\ref{fig-deltas}:
\begin{equation}
\Delta S = S_{\rm out}-S_{\rm in}~.
\end{equation}
The proposal is to weight each vacuum by the entropy increase it admits
\begin{equation}
w_i = \Delta S(i)~.
\end{equation}
Two observers will increase the entropy twice as much as one, so I
have chosen a linear weighting.  (There may be nonlinear effects, for
example a sharp cutoff on the minimum entropy increase required to
have at least one observer; smaller $\Delta S$ would be assigned
weight zero.)
\EPSFIGURE{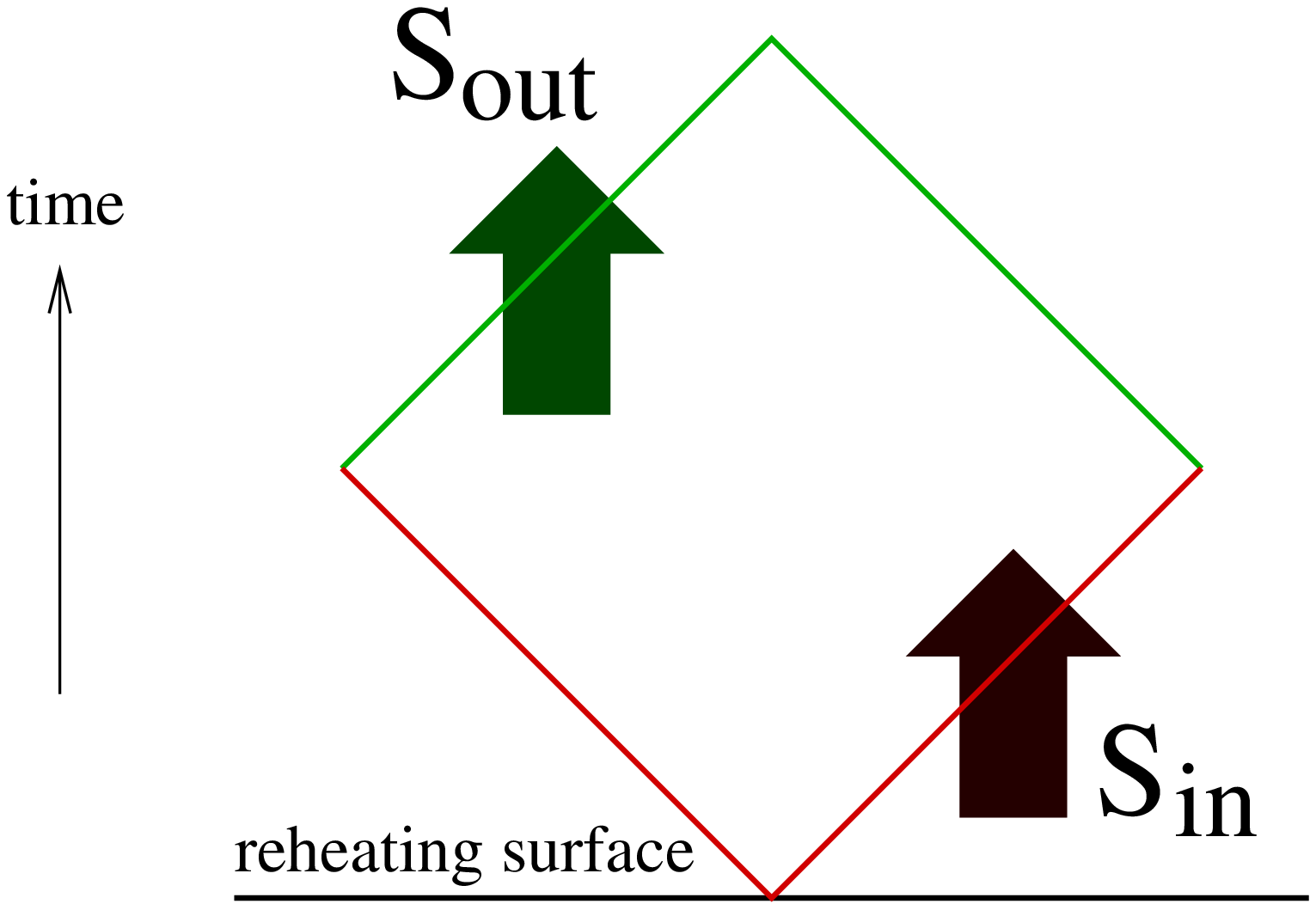,width=.5\textwidth}{\label{fig-deltas}
  Instead of explicit anthropic requirements, a new proposal is to
  weight each vacuum by the amount of entropy, $\Delta S$, produced
  after reheating.  This is the difference between the entropy
  entering the bottom cone of the causal diamond, $S_{\rm in}$, and
  the entropy going out through the top cone, $S_{\rm out}$.}

To be precise, let us take the tip of the bottom cone to lie on the
reheating surface (if there is one; otherwise, no entropy is produced
in any case).  Before this time, the universe is empty, because bubble
formation is strongly suppressed (Sec.~\ref{sec-multiple}).  Only
after reheating will there be matter, and it can organize itself no
faster than at the speed of light.  The tip of the top cone can be
taken to be at a very late time, or even after the vacuum decays; in
the late-time limit the entropy $S_{\rm out}$ will converge quickly.
In a vacuum with positive cosmological constant, the top cone will
coincide with the de~Sitter horizon.

I will not include entropy associated with event horizons.  This would
dominate, particularly through the contribution from the cosmological
horizon in de~Sitter space.  Unlike matter entropy production, it is
not clear how an increase in Bekenstein-Hawking entropy can be related
to the physical process of observation.  However, one could consider
including this contribution for formal simplicity.  In this case, the
argument for prior-based predictions below would need to be augmented
by the extra assumption that in {\em our\/} vacuum, the entropy
produced when black holes are formed, or when they evaporate, is not
related to observers.  The prior-free prediction of $-\log\Lambda$
below would be strengthened, on the other hand, since the horizon
entropy is inversely proportional to $\Lambda$.

Let us compare this ``entropic'' weighting to the anthropic principle.
The latter has been used to predict quantities (such as the
cosmological constant) based on other parameters of our particular
vacuum (such as the time of galaxy formation).  In fact it has {\em
  only\/} been used to make such ``prior-based'' predictions.  Other
examples (some of which happened to be post-dictions) include bounds
on the density contrast $\delta\rho/\rho$~\cite{TegRee97} and on
curvature~\cite{VilWin96,FreKle05}.

In this relatively modest arena, the entropic weighting competes very
well~\cite{BouHar06}.  It turns out that the entropy increase of our
own vacuum is dominated by the photons produced by stars, giving
$\Delta S\approx 10^{85}$.  This means that any variation of
parameters that interferes with star formation will cause $\Delta S$
to drop drastically.  For example, if $\Lambda$ were much larger, no
structure would form, and hence no stars would form, so this
possibility is suppressed by a large drop in $\Delta S$.  In this way,
the entropic weighting reproduces the successes of the anthropic
principle in bounding $\Lambda$, $\delta\rho/\rho$, and curvature in
terms of observed priors.  

This success is remarkable.  The assumptions going into anthropic
arguments are quite specific and detailed.  By contrast, the entropic
weighting is based on a single, simple thermodynamic condition that
observers must satisfy: they must be able to increase the entropy.

In some cases, the entropic weighting will even lead to better
quantitative agreement between predictions and data.  Anthropic
arguments still expect the cosmological constant to be about 100 times
larger than observed~\cite{MarSha97,Vil04,Wei05}.  Large values of
$\Lambda$ are preferred because there are more such vacua, and the
anthropic cutoff is somewhat above the observed value.  In the
entropic weighting, the preference for large $\Lambda$ is weaker: The
overall mass included in the causal diamond scales like
$\Lambda^{-1/2}$.  This shifts the preferred value to smaller
$\Lambda$, in better agreement with observation.

Entropic weighting may allow us to attempt predictions {\em without\/}
priors, a feat thoroughly beyond the ambition of anthropic reasoning.
For example, one might ask where a scale like $10^{-123}$ ultimately
comes from~\cite{Pol06}.  Anthropic arguments only relate the
cosmological constant in our vacuum to the time of galaxy formation
our vacuum.  But in some other vacuum, perhaps stars could have formed
much earlier, allowing the cosmological constant to be much
larger~\cite{Agu01}.

In fact, this is a serious concern.  In the string landscape, many
parameters vary, including $\Lambda$, but also $\delta\rho/\rho$, the
baryon-to-photon ratio, etc.  Taking this into account, is the small
observed value of $\Lambda$ not terribly unlikely after all?  Weinberg
showed only that $\Lambda$ could not be much larger {\em if all other
  parameters are held fixed}.  But they are not, and this may spoil
his explanation of the smallness of the cosmological constant.  (It
cannot spoil his prediction, which, quite sensibly, took observed data
into account.  But it could shift the mystery to questions such as why
$\delta\rho/\rho$ or the baryon-to-photon ratio are so small.)

To address this issue, let us define a weight that depends only on
$\Lambda$, with individual vacua "integrated out":
\begin{equation}
W(\Lambda) d\Lambda= \sum w_i = \sum \Delta S(i)~,
\end{equation}
where $i$ runs over all the vacua with cosmological constant between
$\Lambda$ and $\Lambda+d\Lambda$.  Here $d\Lambda$ should be chosen
large enough for the sum to include a large number of vacua.  Thus
$W(\Lambda)$ is an average weight as a function of $\Lambda$.

The individual weights in this sum will vary hugely.  In fact, I would
expect that $\Delta S(i)$ will typically be quite small.  That is, it
should be atypical to get inflation and reheating, let alone to
dynamically develop complex processes that produce a lot of entropy
after reheating.  But we are interested only in the average of the
weights $w_i$ when summing over a lot of vacua, and in fact we only care
how this average depends on $\Lambda$.  

Let us now make an assumption: suppose that the average is
proportional to (though perhaps much smaller than) the maximum weight
a vacuum can theoretically have, given $\Lambda$.  The entropy
difference cannot be greater than the entropy $S_{\rm out}$.  This is
turn is bounded by the second law of thermodynamics: it must not
exceed the entropy of the cosmological horizon, which is
$3\pi/\Lambda$.  (I am not counting horizon entropy towards $\Delta S$,
since it seems unrelated to the probability of observers, but it can
still be used to bound the entropy produced by matter.)

In fact this bound can be saturated: the total mass inside the horizon
can be up to $\Lambda^{-1/2}$, and the lowest energy quanta one can burn
it into have wavelength $\Lambda^{-1/2}$, so one can produce up to
$1/\Lambda$ quanta.  Of course, one would not expect this extreme limit
to be attained in any significant fraction of vacua (in our own we are
down by $10^{-38}$).   The idea is just that the average weight should
scale in the same way with $\Lambda$ as the maximum weight.  So this
gives
\begin{equation}
W(\Lambda)\propto \Lambda^{-1}~.
\end{equation}

Neglecting for a moment the finiteness of the discretuum density, the
probability for $\Lambda$ to be between $a$ and $b$ will thus be
proportional to $\log a - \log b$.

Now let us assume a discretuum of vacua with roughly even spacing
$1/N$, and $N\approx 10^{500}$.  Thus $\log\Lambda$ will range from
$-500$ to $0$.  According to the above probability, observers should
find themselves at some generic place in this interval, i.e.,
$-\log\Lambda$ should be $O(100)$. 

Clearly, the assumptions going into this argument warrant further
investigation.  Moreover, the result is far less precise than the
Weinberg prediction.  This was to be expected when all recourse to
previously measured quantitites is abandoned.  But it is reassuring
that quite conceivably, the observed value of the cosmological
constant does not become enormously unlikely, even if all other
parameters are allowed to scan; in fact it remains quite typical.

More generally, the argument illustrates that even in the landscape,
we need not give up on predicting observable parameters from the
fundamental theory.  Under the stated assumptions, the order of
magnitude of the logarithm of the size of the universe is related to
the topological complexity of six-dimensional compact manifolds.  This
result is prior free in the sense that it does not use properties of
any particular vacuum, just the structure of the theory.

\acknowledgments I would like to thank many colleagues for useful
discussions, especially A.~Aguirre, B.~Freivogel, L.~Hall, R.~Harnik,
G.~Kribs, A.~Linde, G.~Perez, J.~Polchinski, M.~Porrati, and I.~Yang.

\bibliographystyle{board}
\bibliography{all}
\end{document}